\documentclass[sn-mathphys-num]{sn-jnl}% Math and Physical Sciences Numbered Reference Style
\usepackage{graphicx}%
\usepackage{multirow}%
\usepackage{amsmath,amssymb,amsfonts}%
\usepackage{amsthm}%
\usepackage{mathrsfs}%
\usepackage[title]{appendix}%
\usepackage{xcolor}%
\usepackage{textcomp}%
\usepackage{manyfoot}%
\usepackage{booktabs}%
\usepackage{algorithm}%
\usepackage{algorithmicx}%
\usepackage{algpseudocode}%
\usepackage{listings}%

\newcommand{\TrtHz}{$\text{T}/\sqrt{\text{Hz}}$}
\theoremstyle{thmstyleone}%
\theoremstyle{thmstyletwo}%

\theoremstyle{thmstylethree}%

\begin{document}

\title[Article Title]{Silicon-photonic optomechanical magnetometer}


\author[1,2]{\fnm{Fernando} \sur{Gottardo}}

\author[1,2,3]{\fnm{Benjamin J.} \sur{Carey}}

\author[1,2,3]{\fnm{Nathaniel} \sur{Bawden}}

\author[1,2]{\fnm{Glen I.} \sur{Harris}}

\author[1,2]{\fnm{Hamish} \sur{Greenall}}

\author[1,2]{\fnm{Erick} \sur{Romero}}

\author[4]{\fnm{Douglas} \sur{Bulla}}

\author[5,6]{\fnm{James S.} \sur{Bennett}}

\author[4]{\fnm{Scott} \sur{Foster}}

\author*[1,2,3]{\fnm{Warwick P.} \sur{Bowen}}\email{w.bowen@uq.edu.au}

\affil[1]{\orgdiv{School of Mathematics and Physics}, \orgname{The University of Queensland}, \orgaddress{\city{St Lucia}, \postcode{4067}, \state{Queensland}, \country{Australia}}}

\affil[2]{\orgname{ARC Centre of Excellence for Engineered Quantum Systems}, \orgaddress{\city{St Lucia}, \postcode{4067}, \state{Queensland}, \country{Australia}}}

\affil[3]{\orgname{ARC Centre of Excellence in Quantum Biotechnology}, \orgaddress{\city{St Lucia}, \postcode{4067}, \state{Queensland}, \country{Australia}}}

\affil[4]{\orgname{Australian Government Department of Defence Science and Technology}, \orgaddress{\city{Edinburgh}, \postcode{5111}, \state{South Australia}, \country{Australia}}}

\affil[5]{\orgdiv{Centre for Quantum Dynamics}, \orgname{Griffith University}, \orgaddress{\city{Nathan}, \postcode{4111}, \state{Queensland}, \country{Australia}}}

\affil[5]{\orgdiv{School of Mathematical Sciences}, \orgname{Queensland University of Technology}, \orgaddress{\city{Gardens Point}, \postcode{4000}, \state{Queensland}, \country{Australia}}}


\abstract{Optomechanical sensors enable exquisitely sensitive force measurements, with emerging applications across  quantum technologies, standards, fundamental science, and engineering. Magnetometry is among the most promising applications, where chip-scale optomechanical sensors offer high sensitivity without the cryogenics or magnetic shielding required by competing technologies. However, lack of compatibility with integrated photonics and electronics has posed a major barrier. Here we introduce silicon-on-insulator optomechanical magnetometers to address this barrier. A new post-release lithography process enables high-quality metallisation of released mechanical structures, overcoming the incompatibility between silicon-on-insulator fabrication and functional magnetic films. This allows us to employ photonic-crystal cavities that enhance motion-to-optical signal transduction by over an order of magnitude. The resulting devices achieve magnetic field sensitivity of 800\,p\TrtHz, three orders of magnitude beyond previous waveguide-integrated designs. The advances we report provide a path towards high-performance, room temperature and chip-integrated  magnetometers for applications ranging from biomedical imaging and navigation to resource exploration.

}




\maketitle

% \section{Introduction}\label{sec1}

{Originally pioneered to enable gravitational wave detection~\cite{abramovici1992ligo}, optomechanical sensors harness optical and mechanical resonances to achieve exquisitely sensitive measurements~\cite{li2021cavity}. Recent years have seen a rapid acceleration in  chip-scale optomechanical sensing, driven by applications in quantum computation and communications~\cite{barzanjeh2022optomechanics}, %standards~\cite{???}, and industry~\cite{???}, 
and in fundamental science such as dark matter detection~\cite{brady2023entanglement, baker2024optomechanical} and the foundations of quantum mechanics~\cite{aspelmeyer2014cavity}. These sensors are now able to measure forces and displacements with attonewton~\cite{gavartin2012hybrid, mason2019continuous} and attometre~\cite{schliesser2008high} sensitivity, respectively, and allow ultrasound and acceleration measurements with otherwise inaccessible precision~\cite{basiri2019precision, cao2025integrated, bawden2025precision}. They are even able to resolve quantum zero-point motion~\cite{huang2024room}, and have enabled a new class of quantum-calibrated temperature standards~\cite{Purdy2017}.

Magnetometry is a particularly important application area, with potential impact across medical imaging~\cite{brookes2022magnetoencephalography, murzin2020ultrasensitive, borna2020non}, navigation~\cite{Bennett2021, connerney2015maven}, chemical analysis~\cite{savukov2005nmr, devience2015nanoscale}, and geological surveying~\cite{stolz2022squids, lu2023recent, stele2023drone}. Unlike other precision magnetometers~\cite{storm2017ultra,Savukov2013}, optomechanical magnetometers require no cryogenics or magnetic shielding~\cite{Forstner2012, Forstner2014},
%generally require either cryogenics~\cite{storm2017ultra} or magnetic shielding~\cite{Savukov2013}. Optomechanical magnetometers~\cite{Forstner2012, Forstner2014} promise to overcome this challenge, 
providing high magnetic field sensitivity at room temperature and in background fields as large as that of the Earth~\cite{Bennett2021}. Recent advances have reached hundreds-of-femtotesla~\cite{xu2024subpicotesla} sensitivity, with sensitivity now rivaling that of state-of-the-art cryogenic magnetometers of similar size~\cite{stolz2022squids, Li:20, hu2024picotesla}. However, previous optomechanical magnetometers~\cite{Forstner2012, Forstner2014, Li:18Quant, Li:20, gotardo2023waveguide, hu2024picotesla} have had limited compatibility with integrated photonics or electronics, constraining both performance and applicability. We overcome this barrier, developing an optomechanical magnetometer based on silicon-on-insulator (SOI), an 
%SOI is compatible with silicon photonics and CMOS electronics~\cite{shekhar2024roadmapping}, making it the 
industry-standard for scalable photonic integration~\cite{shekhar2024roadmapping,atabaki2018integrating}. 

Previous work has found that standard SOI fabrication processes degrade functional metallic films~\cite{dai2021recent}, such as those used in optomechanical magnetometers~\cite{hu2024picotesla, gotardo2023waveguide}. We address this significant challenge, developing a new post-release lithography technique which enables high-quality metallisation of delicate mechanical elements and is readily transferable to other microsystems requiring functional films on released structures, such as MEMs magnetometers \cite{ramasamy2011magneto, herrera2016recent}, magnetoelectric and magnetoacoustic transducers~\cite{luo2024magnetoelectric, dreher2012surface}, spin-filters \cite{tereshchenko2025direct}, bolometers \cite{liu2020design, dao2019mems}, and quantum information devices \cite{pirkkalainen2013hybrid, o2010quantum}. 

The use of post-release lithography allows us to create complex functional devices on SOI. Going beyond previous whispering-gallery-based optomechanical magnetometers~\cite{Forstner2012, Forstner2014, Li:18Quant, Li:20, gotardo2023waveguide, hu2024picotesla}, it allows optomechanical magnetometers that co-integrate magneto-mechanical and photonic structures, independently optimized for magnetic coupling and optical readout, and with photonic crystal resonators for readout. This increases the transfer of magnetic-field-induced motion into optical signal by a factor of 27. %{\bf WPB later on you quote Gom, and the factor seems more like 10}. This effectively eliminates optical noise across the full measured spectrum. 
%co-integrate magneto-mechanical and photonic structures, creating waveguide-coupled optomechanical magnetometers that are independently optimized for magnetic coupling and optical readout (Figure~\ref{fig:working principle NB}). Using an engineered photonic crystal cavity for readout increases the transfer of motion into optical signal by a factor of 100, effectively eliminating optical noise across the full measured spectrum. 
The resulting sensor achieves a peak sensitivity of 800\,p\TrtHz, more than three orders of magnitude better than previous waveguide-integrated devices \cite{gotardo2023waveguide}.

Together, these advances pave the way for scalable chip-scale optomechanical magnetometers with potential applications ranging from biomedical imaging to geological surveying.
}
%They use magnetostriction to convert magnetic fields into mechanical motion, and an optical cavity to observe this motion with exquisite precision~\cite{Forstner2012, Forstner2014}. This 

%Here, they enable precision magnetometry without the cryogenic systems or magnetic shields that are otherwise needed [???]. This could revolutionise applications ranging from magnetic brain imaging [???], to magnetic navigation and…
 
%However, optomechanical magnetometers to-date have not been compatible with integrated photonics or electronics.  This both impedes wide-spread applications and limits design choices, and therefore magnetometer performance. Here, we resolve this challenge…
 
%To realize their full potential, scalable silicon-chip integration with other optical and electronic components is essential. However, previous on-chip optomechanical magnetometers have used a silica device layer~\cite{Forstner2012, gotardo2023waveguide,Forstner2014, zhu2017polymer, Li2018}. Silica’s low refractive index precludes scalable integration~\cite{soref2007past}. It also inhibits the use of complex photonic structures. This results in relatively weak transfer of mechanical motion into optical signal, with sensitivity limited by optical noise at most frequencies~\cite{hu2024picotesla}, and substantially degraded sensitivity in waveguide-coupled devices~\cite{gotardo2023waveguide}.

\section{Fabricating functional suspended devices on silicon-on-insulator}
\label{Sec:Fab}

\begin{figure}[htbp]
    \centering
    \includegraphics{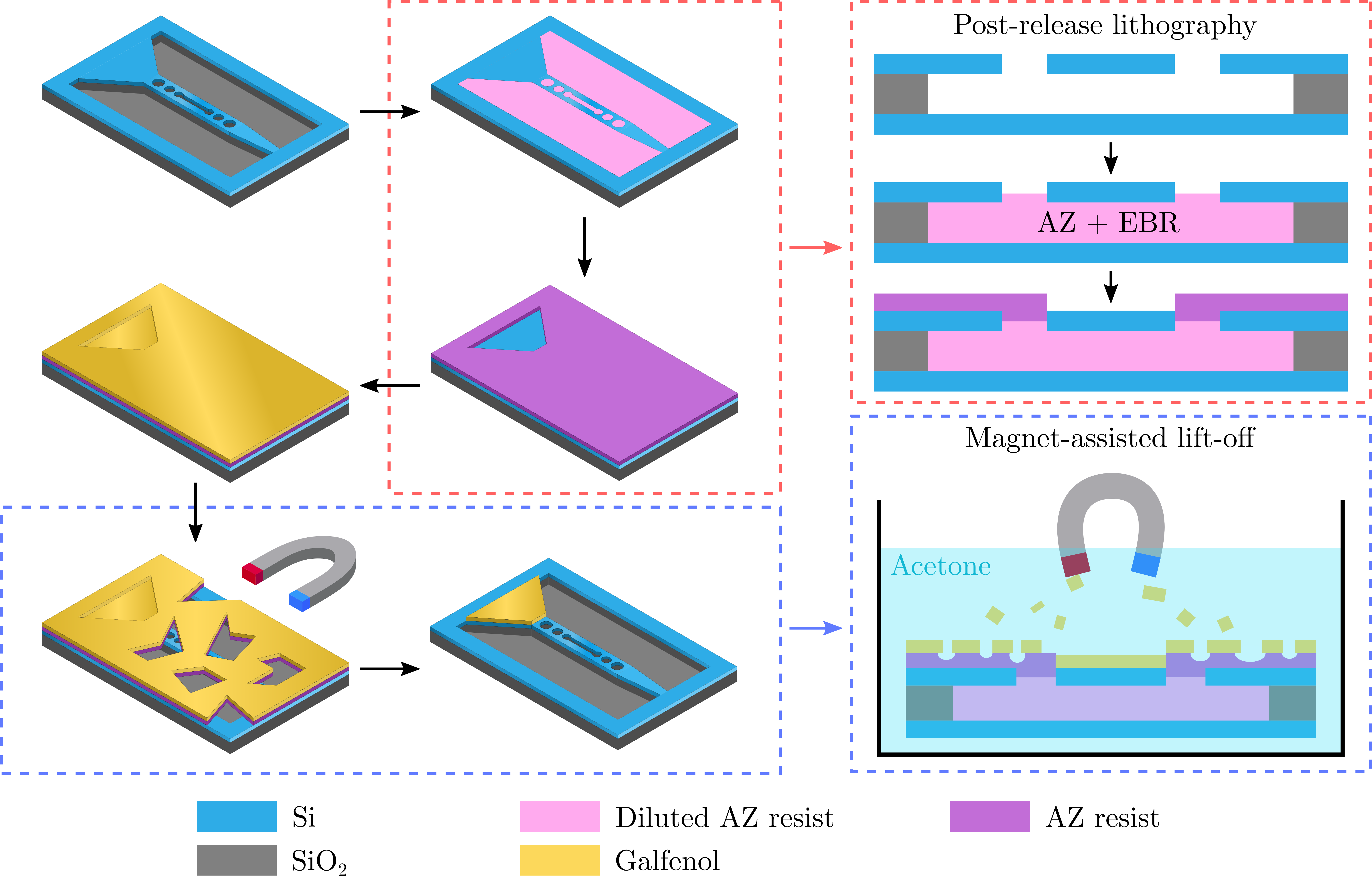}
    \caption{Overview of the post-release fabrication process. A freestanding optomechanical device is flooded with diluted resist solution (AZ + EBR) prior to photolithography and galfenol deposition. The excess galfenol and structure supporting resist are removed via a magnetically-assisted lift-off process, followed by critical point drying, resulting in the final device.}
    \label{fig:3D recipe}
\end{figure}

%{\color{blue} Previous on-chip optomechanical magnetometers >>> silica on silicon. Advantage: highly selective and gentle xenon difluoride gas etch available for release. Disadvantages: low refractive index of silica presents major barrier to integration and precludes the use of complex geometries. As a consequence, all prior on-chip optomechanical magnetometers have employed ring-shaped whispering gallery mode resonators that exhibit weak optomechanical coupling and present challenges for optimising transduction of magnetic strain into mechanical motion.

%Silicon-on-insulator (SOI), release step much harsher and less selective (see references below), employs hydrofluric acid, know to degrade magnetostrictive materials -- and we verify this.

%To address this, and enable broader functionalisation of SOI devices, without interference from HF etching, we develop...
%}

{Optomechanical magnetometers function by detecting minute motions caused by magnetic-field-induced stress, or magnetostriction~\cite{Forstner2012}. They achieve high sensitivity by using optical cavities to precisely measure the magnetostrictive response of compliant suspended structures, whose motion is enhanced by mechanical resonances~\cite{li2021cavity}. All previous on-chip optomechanical magnetometers have used a silica device layer~\cite{Forstner2012, Forstner2014, Li:18Quant, Li:20, gotardo2023waveguide, hu2024picotesla}. This has significant advantages; providing access to ultrahigh quality optical resonances~\cite{armani2003ultra, lee2012chemically} and allowing  gentle, selective release of fragile suspended structures via xenon difluoride vapor phase etching~\cite{yang2018bridging, bekker2018free}. However, silica's low refractive index limits compatibility with integrated electronics and photonics, and makes it difficult to engineer complex and compact structures~\cite{su2023scalability}. 
%Indeed, attempts to integration with on-chip waveguides alone have resulted sensitivity far below that of non-integrated devices~\cite{???}.
%The low refractive index 
Indeed, all previous on-chip optomechanical magnetometers have employed simple circular whispering gallery mode cavities~\cite{Forstner2012, Forstner2014, Li:18Quant, Li:20, gotardo2023waveguide, hu2024picotesla}, with far weaker signal transduction from mechanical to optical domains than can be achieved in more complex devices~\cite{eichenfield2009picogram, Winger_2011}.

To enable full integration and provide access to the rich spectrum of optomechanical device architectures that have developed in recent years~\cite{engelsen2024ultrahigh, sementilli2022nanomechanical, ghadimi2018elastic, fedorov2019generalized, tsaturyan2017ultracoherent, bereyhi2022hierarchical, beccari2022strained}
% aspelmeyer2014cavity -- wrong reference. This needs to be several science and nature papers on things like disispation dilution, soft clamping... you could inlcude leos review. Please be careful about references - check others as well. We wouldn't get away with citing a 2014 review paper as "recent" developments.
, it is necessary to translate optomechanical magnetometry onto a higher refractive index device layer. Silicon-on-insulator (SOI) -- consisting of a silicon device layer atop a silica sacrificial layer -- is particularly attractive, since it is a primary industry standard for photonic integration~\cite{sun2015single, stojanovic2018monolithic}. %I would add the reference to SOI integration in Chris/Tihan's full FSR tuning paper (nature or science paper).
%However, this is challenging because under-etching of the SOI silica sacrificial layer to release functional devices generally requires hydrofluoric acid etching~\cite{???}, which is known to degrade a range of metal alloys~\cite{dai2021recent}. 
To release functional devices, it is necessary to under-etch the sacrificial layer. In SOI, this generally requires vapor phase hydrofluoric acid etching~\cite{lee1997dry}, which is known to degrade a range of metal alloys~\cite{dai2021recent}. 
%this is challenging because the release of functional devices involves under-etching of the sacrificial layer. This generally requires vapor phase hydrofluoric acid etching~\cite{???}, which is known to degrade a range of metal alloys~\cite{dai2021recent}. 
We attempted vapor phase hydrofluoric acid etching of devices coated with the magnetostrictive material, galfenol (Fe$_\text{81}$Ga$_\text{19}$)~\cite{Nivedita2017galf, adolphi2010improvement, greenall2024QPM}. Even 
%Consistent with these observations, we found that  
with a protective tantalum capping layer, fluorine penetration during etching eliminated any measurable response (see Sec. S.1). An aluminium capping layer was also tested, but did not exhibit a magnetic response due to HF attacking the galfenol through the exposed sidewall. The  galfenol could potentially be fully encapsulated using a second lithography step to define a region slightly larger than the galfenol and depositing an appropriately thick aluminium layer. However, this would restrict the sensors motion, and add a second round of lithography, deposition, and liftoff. % This would also require a gap between the galfenol and the edge of the cantilever to allow encapsulation, which would reduce the area of galfenol that could be used and reduce the magnetic response. 
%{\bf WPB: wouldn't this also reduce the area of galfenol that could be used, reducing the magnetic response? I.e., you need a gap between the galfenol and the edge of the cantilever to allow encapsulation.}. %{\bf WPB: did we also try with terfenol? Is there anything else we tried that's worth mentioning?}.
%from the magnetostrictive material, galfenol, that we used to for device functionalisation  

To remove degradation from the release step, we developed a new post-release lithography and deposition process that is robust and broadly generalisable (Fig.~\ref{fig:3D recipe}). In the usual fabrication process, because of the extreme fragility of the suspended structures, the release step is performed last -- after deposition of the magnetostrictive material, and exposing it to risk of degradation~\cite{li2021cavity}. Our new process flips this order, depositing the magnetostrictive material \textit{after} release. This means that the thin-film is deposited at the end of the process, avoiding any deleterious effects from the other fabrication steps. The major challenge then is to protect the fragile released structures during lithography and deposition. This hurdle is specific to technologies that require thin device layers -- in our case hundreds of nanometers thick, to support single-mode silicon photonics~\cite{atabaki2018integrating}. In contrast, MEMs typically utilise a much thicker device layer, up to 10\,\textmu m~\cite{ilgaz2024miniaturized}, greatly increasing robustness and resilience to subsequent fabrication procedures.

{A critical step to achieve post-release lithography was the development of a method to support the released devices, embedding them in partially solidified photoresist (Fig.~\ref{fig:3D recipe}). For this step, the photoresist was diluted with edge bead removal solvent with an equal volume of edge bead removal solvent, reducing its viscosity by a factor of five. Combined with the factor of four reduction in spin speed, and including the change in resist thickness due to the modified viscosity, this results in a factor of twenty lower shear load during spinning compared to undiluted resist. The diluted resist acts as an elastic support for the suspended device and can be modeled as an elastic Winkler Foundation~\cite{hetenyi1946beams}. Within this model, the resist exerts a distributed restoring force, with characteristic length $\lambda^{-1} = (\frac{4EI}{k})^{1/4} < 1$\,\textmu m, where $I = \frac{w t^3}{12}$, $k \approx \frac{E_\textrm{resist}}{h}$, $E$ is the Young’s modulus of the suspended structure with width $w$ and thickness $t$, and $E_\textrm{resist}$ is the Young’s modulus of the resist with thickness $h$. Because the length of the suspended structure is much larger than the characteristic length, the foundation exerts strong restraint on the beam and constrains out-of-plane motion, suppressing global bending and thereby protecting the delicate released structures during subsequent fabrication steps. We found that this post-release lithography process did not harm the released structures, provided sufficient support and allowed easy removal during the final liftoff of the excess magnetostrictive material. Attempts using undiluted photoresist, with its relatively high viscosity and requiring typical spin coating speeds, broke the structures.}

%For this step, the photoresist was diluted with edge bead removal solvent to reduce its viscosity and allow gentle spin coating at a low speed of 1000 RPM. With an empirically determined dilution ratio of equal parts photoresist and solvent, we found that this process did not harm the released structures, provided sufficient support during the subsequent fabrication procedures, and allowed easy removal during the final liftoff of the excess magnetostrictive material. Attempts using undiluted photoresist, with its relatively high viscosity and requiring typical spin coating speeds, broke the structures.

%{\bf WPB: we need more substance here (or when discussing this part of the process later). The impression can't be of a trivial extension of existing approaches. Perhaps brainstorm with Fernanodo and Elliot.}
%{\bf WPB: we need to add more details here -- why was this difficult/innovative? what did we learn?... We do need to add more substance here.}

The fabrication process, in brief, is as follows (see Methods for full details): an SOI chip, comprising a silicon device layer of 215\,nm approximate thickness atop a 3\,\textmu m thick sacrificial silica layer, is patterned using electron beam lithography (EBL) and etched with reactive ion etching. The devices are then released using HF vapor phase etching. After release, a low-viscosity resist is spin-coated to fill the etched regions and support the suspended elements. This support layer is cured and reinforced with a second coating, allowing standard photolithography to define the galfenol deposition area without collapsing the devices (Fig.~\ref{fig:3D recipe}). A 600\,nm galfenol layer is then deposited by DC sputtering.

%involves patterning and functional material deposition, followed by HF etching of the silica to release the patterned structures -- the stage in the process in which the degradation occurs. Our new process flips 

}

Following deposition, excess galfenol is removed via a lift-off process. However, typical lift-off processes involving sonication in solvent were found to damage the released structures. To address this, we devised an alternative, magnetically assisted lift-off process (Fig.~\ref{fig:3D recipe}). Here, the chip is submerged in acetone and a neodymium permanent magnet is placed approximately 5\,mm above. This magnet promotes the removal of galfenol and prevents redeposition on the chip. At the same time, the acetone removes the semi-solidified resist supporting the structure. The chips are then dried with critical-point carbon dioxide. %\bf WPB: I feel like there's a point missing here... ...how did we remove the semi-solidified resist, what did we learn from this? This also isn't clear in the figure caption.

%{\color{blue} Mention how general the approach is: metalisation after lift-off, immune to harsh etching required for lift-off}

We characterised the magnetostrictive properties of the galfenol films after the fabrication process using optical profilometry \cite{greenall2024QPM}. In contrast to the conventional method, where no response was observed, the films
%The galfenol films used 
were found to present piezo-magnetic stress and strain coupling coefficients %($\alpha_{33}$ \& $d_{33}$ respectively) 
of $\alpha_{33}=2\times10^8$\,Pa/T and $d_{33}=3.2$\,nm/A, respectively
%determined using optical profilometry \cite{greenall2024QPM} 
(these values were used in the modeling shown in Fig.~\ref{fig:Sensitivity}d and Sec.~S.3.3). 
The piezomagnetic coefficients of the thin film determine the efficacy of transduction of magnetic fields to mechanical motion via applied stress and strain, respectively. The measured values are comparable to other thin-film galfenol measurements \cite{greenall2024QPM, shi2019study, García-Arribas2021}, with the highest reported strain coupling coefficient being 16\,nm/A \cite{García-Arribas2021}.

%\textcolor{violet}{To enable this, we developed fabrication techniques including post-release lithography and magnetically assisted lift-off. While motivated by the demands of magnetostrictive sensing, these methods could also support scalable manufacture of a broader class of microsystems that rely on suspended metalized elements, such as accelerometers, electrostatically actuated switches, resonant pressure sensors, and micro-mirrors for optical beam steering.}

\section{Optomechanical magnetometry on silicon-on-insulator}
\label{sec:design}

\begin{figure}[htbp]
    \centering
    \includegraphics{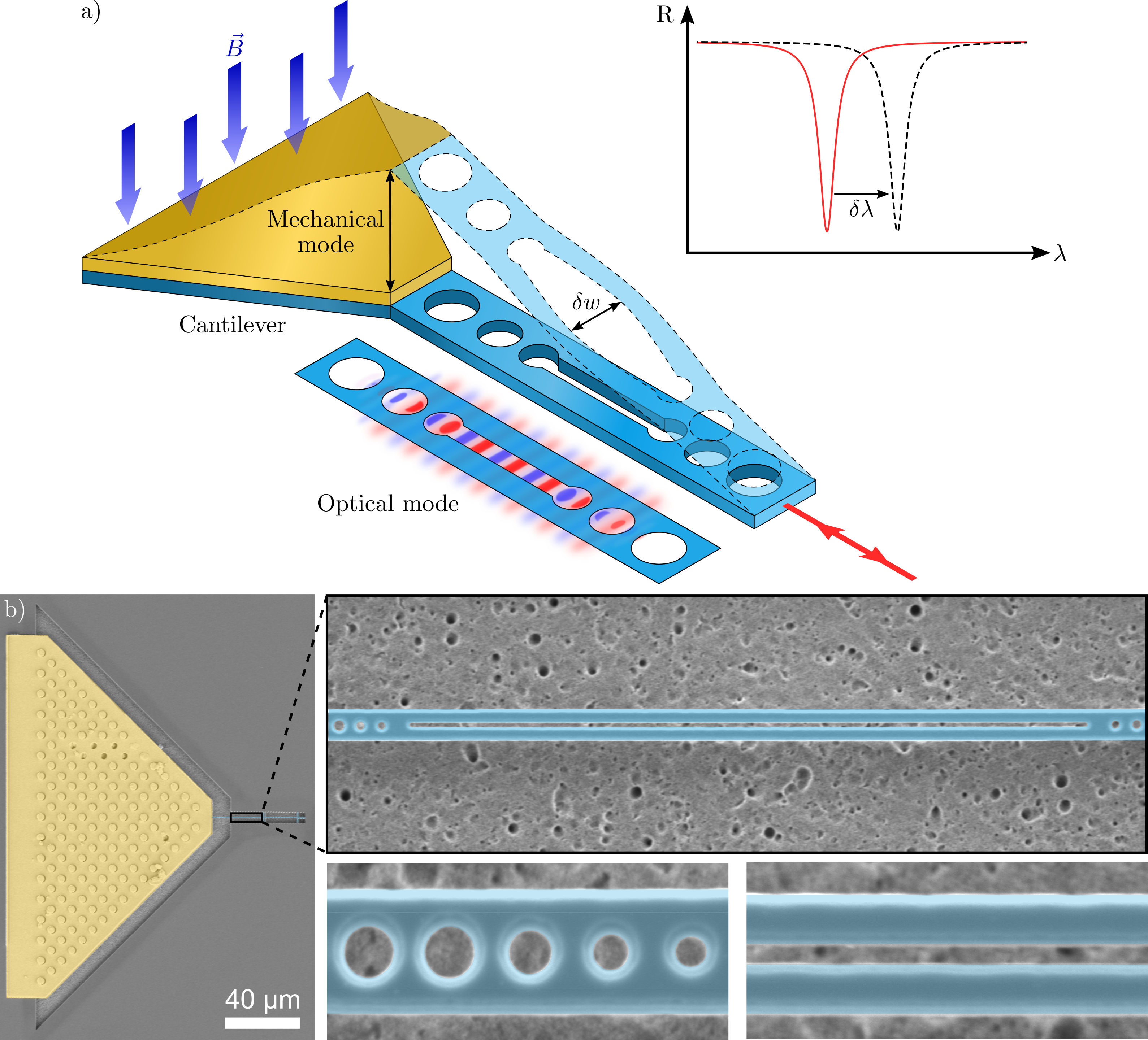}
    \caption{a) Schematic diagram of the device design and working principle. A magnetic field ($\vec{B}$) induces deformation of a galfenol (yellow) platform, which pulls on a silicon (blue) slot cavity. The resulting in-plane deformation of the cavity ($\delta w$) induces a change in its optical resonance wavelength ($\delta \lambda$). This can be detected by probing the reflected light. The red arrows indicate coupling of the optical mode to an on-chip waveguide. b) False color SEM image of a magnetometer with close-up images of the slot cavity. Silicon, blue; galfenol, yellow.}
    \label{fig:working principle NB}
\end{figure}

\begin{figure}[htbp]
    \centering
    \includegraphics{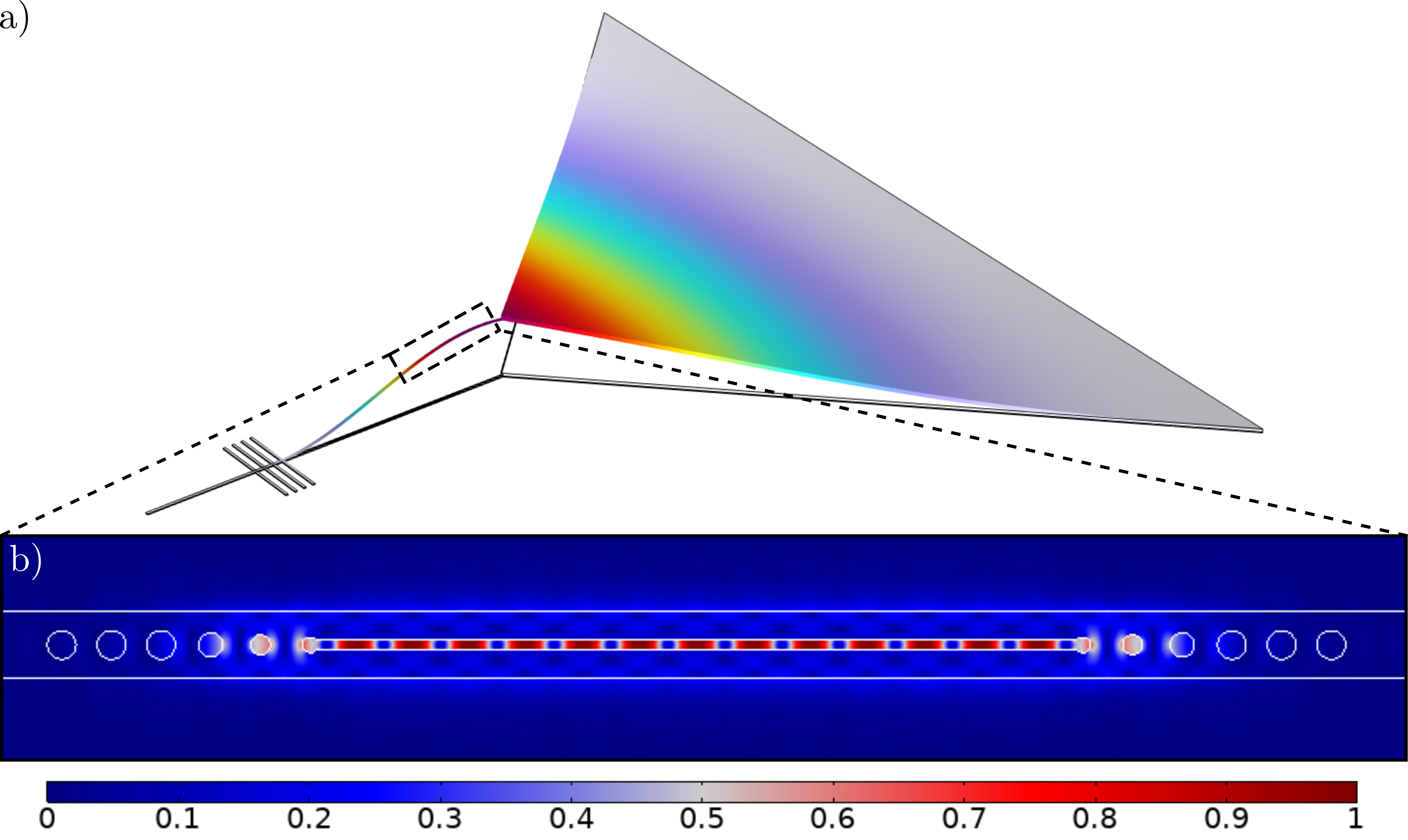}
    \caption{Finite element method modeling with COMSOL. a) Modeshape of the fundamental out-of-plane vibrational mode. b) Intensity distribution of the 1D photonic slot cavity's optical mode (normalised by the peak intensity).}
    \label{fig:modeling1}
\end{figure}

{The availability of post-release lithography and deposition allowed fabrication of optomechanical magnetometers on SOI, exploiting the high refractive index silicon device layer to enable complex designs. The magnetometer design is shown in Fig.~\ref{fig:working principle NB}, along with a scanning electron micrograph of a fabricated device. The magnetometers consist of a photonic crystal cavity, connected to a separate magnetostrictive actuator as well as an on-chip waveguide for optical excitation and read-out. Light is coupled to and from the waveguide using a single-sided tapered fiber, which can be bonded to the waveguide using ultraviolet-curing adhesive~\cite{wasserman2022cryogenic} (See Section~S.2).%Thus, our device is a fully integrated optomechanical magnetometer.

The physical separation of cavity and magnetic actuator allows each to be independently optimised. The actuator consists of a triangular mechanical cantilever with an out-of-plane mechanical `flapping mode' resonance that is coated with a film of galfenol. Applying a magnetic field produces strain within the galfenol film, which in turn drives mechanical motion of the cantilever. The tip of the cantilever is connected to one end of the photonic crystal cavity, 
%The actuator is physically connected to an optical resonator, 
facilitating the conversion of mechanical motion  into measurable optical frequency shifts.

%We employ the new post-release lithography and deposition process to 
%%enables the fabrication complex silicon-photonic devices to be fabricated. Here, we use it to 
%develop a new class of optomechanical magnetometers. 
%These magnetometers comprise a photonic crystal cavity that is designed convert mechanical motion into optical signals with high efficiency, connected to a separate magnetostrictive actuator optimised to transduce magnetic signals into mechanical motion, as well as to an on-chip waveguide for optical excitation and read-out.
%, and a separate magnetostrictive actuator optimised for transduction of magnetic signals into mechanical motion.

%The optical mode is interrogated via an integrated Si coupling waveguide at the end of the cavity. 

%, and a photonic resonator that is affixed to the unclamped end of the cantilever. These facilitate the magneto-mechanical transduction and optomechanical readout, respectively. By separating the design of the mechanical resonator from the optical cavity, the magnetic actuation and optical readout can be optimized independently. The cantilever is coated with a thin film of the magnetostrictive material galfenol (Fe$_\text{81}$Ga$_\text{19}$) \cite{Nivedita2017galf, adolphi2010improvement, greenall2024QPM}. 

The photonic crystal cavity is defined by a silicon waveguide with photonic crystal (PhC) mirrors at either end~\cite{velha2007ultra}. We etch a slot along the center of the waveguide.  
%, modulating the optical resonance frequency.
This introduces a set of gap-modes~\cite{almeida2004guiding}, the optical fields of which are strongly concentrated within the slot (Fig.~\ref{fig:modeling1}b). The resonance frequencies of these modes are exquisitely sensitive to changes in the slot width,
%such When the cantilever pulls on the cavity it induces a change in the width of this slot.
amplifying the effect of mechanical motion on the optical resonance frequency. Moreover, the presence of the slot greatly increases the compliance of the cavity, increasing the mechanical displacements induced by magnetostrictive forces. Together, this significantly enhances the conversion from magnetic field to optical signal.

%to optimise the conversion of magnetostrictive strain into measurable optical frequency shifts. The optical field is concentrated into a gap-mode within this slot~\cite{li2021silicon}. 

%Compared to other photonic crystal designs~\cite{???}, and to previous whispering gallery mode-based magnetometers~\cite{???}, the use of a photonic crystal slot cavity has significant advantages. First, the resonance frequency of the gap-mode is exquisitely sensitive to changes in the slot width, amplifying the effect of mechanical motion on the optical resonance frequency. Second, the presence of the slot greatly increases the compliance of the cavity, so that magnetostrictive forces cause larger mechanical displacements. }

We use Finite Element Method (FEM) modeling (COMSOL Multiphysics) to optimise the design of the magnetometer.
%was used to predict both the mechanical and optical behavior of the system. 
Fig.~\ref{fig:modeling1}a depicts the fundamental mechanical mode of the magnetic actuator, with a resonance frequency of 250\,kHz. %this relatively high frequency (given the dimensions) is due to residual stress in the magnetostrictive film. 
It is apparent that the displacement of the actuator is primarily out of plane. However, this causes an elongation of the highly mechanically compliant photonic cavity, which in turn drives a change in the slot width. The optical eigenmode of the slot cavity is depicted in Fig.~\ref{fig:modeling1}b. This confirms that the majority of the electric field is tightly confined within the air gap of the slot. 

% The shape of the slot is engineered to efficiently transfer the mechanical displacement of the cantilever into a change in slot width. 
% We found that including circular holes connected to both ends of the slot improved the transfer of lateral strain into a transverse change in slot width  {\bf WPB: correct?}. 
Employing a longer slot amplifies the change in slot width due to applied magnetic fields by the increased mechanical compliance. However, it also increases the slots thermomechanical motion, obscuring the motion of the galfenol cantilever and reducing sensitivity. The transfer efficiency from cantilever motion to slot motion  must therefore be balanced against slot thermomechanical noise to optimise  sensitivity. We found that a 7\,\textmu m slot length to be suitable for this purpose. 
%This improves with the length of the slot but also increases its thermomechanical motion. 
The conversion of mechanical motion to optical frequency shift is quantified by the optomechanical coupling rate, $G_\text{OM} = \partial\omega/\partial x$, where $x$ is the slot width.
%, and is independent of the slot length. %In a slot photonic crystal, the optomechanical coupling  rate increases as the slot width decreases, but saturates to a constant value once the width is far-sub-wavelength {\bf WPB: is this statement correct?}
We modeled this dependency for our slotted photonic crystal designs (Sec.~S.3.1), finding that the optomechanical coupling rate increases as the width of the slot decreases, and eventually rolls off at small slot widths. % {\bf WPB: the connection here isn't good. (1) based on the previous sentence the "consequence" should be that we use zero width. (2) there's no discussion of length at all in the previous sentence. Needs more elaboration.} 
We chose a width of 100\,nm, providing a high optomechanical coupling rate, while also allowing reliable fabrication without risk of the slot collapsing or not being fully released.
%balancing high optomechanical coupling rate with thermomechanical motion and reliability of fabrication. 
%{\color{blue} \bf WPB: it's not clear from this why we chose the length as we did. Can you clarify this? It may be that the choice of length would be better stated earlier just after mentioning the increased thermomechanical noise? I woudl say that the compromise there isn't so clear. All equal, I think you'd just make the length as long as you could, so that you're thermomechanical noise limited everywhere (this will give optimum sensitivity. I guess the compromise here is really that it's difficult to fabricate very long slots without them breaking, collapsing, or being too fragile to operate with.)}
%is beneficial for optomechanical sensing as it increases the mechanical signal in comparison to the optical noise-floor. 
%A gap of 100\,nm was chosen for our devices to ensure the repeatability of fabrication. 
This separation yields a coupling rate of 99.8\,GHz/nm, comparable to the highest reported for optomechanical cavities \cite{navarathna2024silicon, Winger_2011, Xia_Optomech_review_2020, woolf2013optomechanical} and a factor of twenty-seven higher than the best previous chip-scale optomechanical magnetometers~\cite{hu2024picotesla}.  %which boast coupling rates of 3.6\,GHz/nm \cite{hu2024picotesla}
% {\bf WPB: we're 27 times better, but say 20 in the intro}. %The optomechanical coupling corresponding to changes in the cavity length is orders of magnitude lower, so we therefore neglect it.

%{\bf WPB: can we add specific insights? Also, how did we choose its length? I guess longer is just better in principle, but theres a limit to length that's possible in practice. Are ours the longest slot photonic crystals ever made? }. {\bf WPB: are we able to quantify the net effect compared with using a standard suspended photonic crystal? I.e. How much larger is the frequency shift of the slot cavity per unit strain.}
}

%{\color{blue} A fabricated device is shown in Fig.~\ref{fig:SEM}.}

% \begin{figure}[htbp]
%     \centering
%     \includegraphics{Figures/SEM.png}
%     \caption{False color SEM of the magnetometer with close-up images of the slot cavity. Silica, gray; silicon, green; galfenol, blue.}
%     \label{fig:SEM}
% \end{figure}

\section{Waveguide-integrated magnetometer with sub-nanotesla sensitivity}
\label{sec:perf}

\begin{figure}[htbp]
    \centering
    \includegraphics{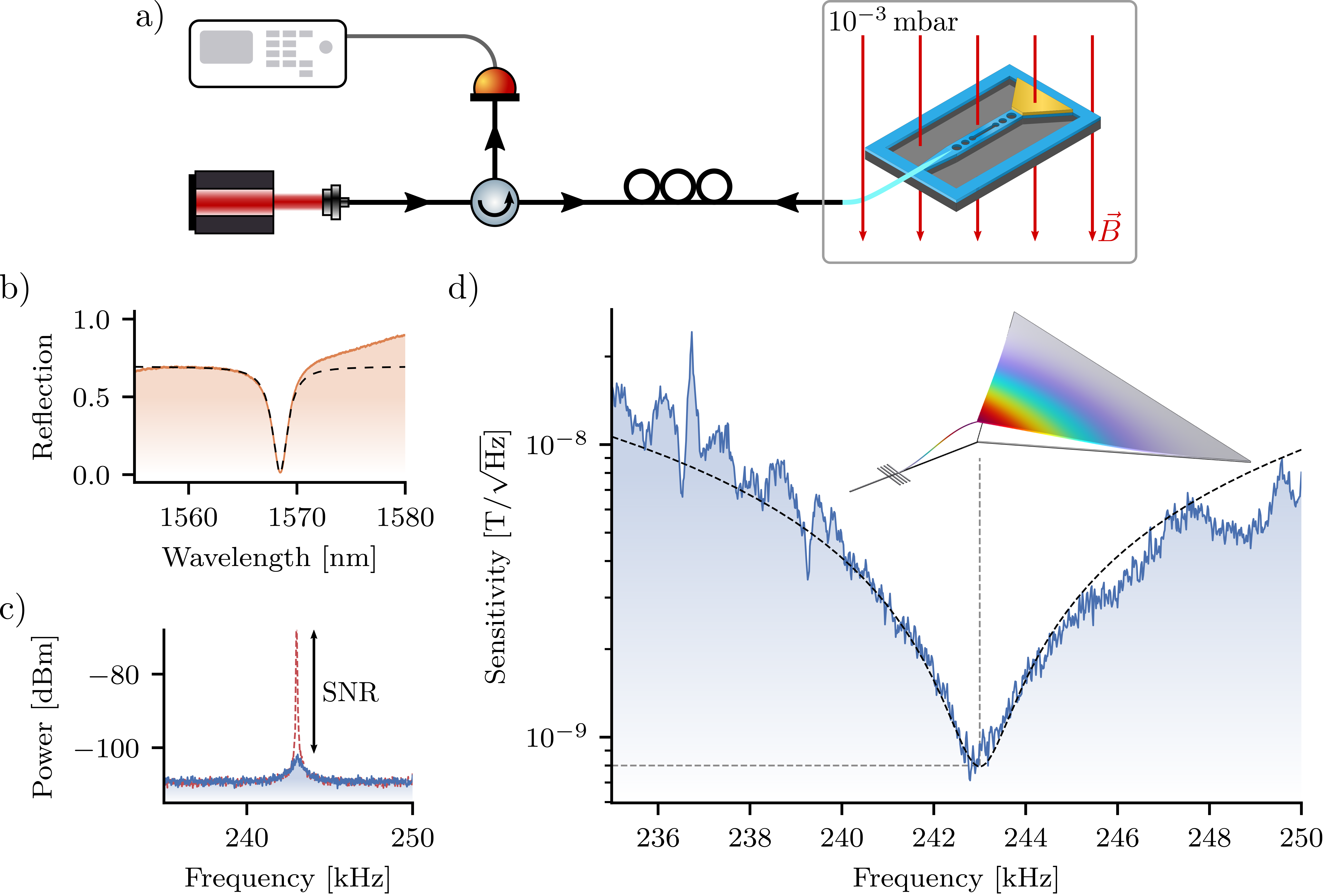}
    \caption{Magnetometer characterization. a) Experimental setup for sensitivity measurements. b) Optical spectrum reflected from the slot cavity. The Lorentzian fit (dashed) gives an optical quality factor of 930. c) Power spectrum showing the mechanical mode and the signal from an applied field measured with a 30\,Hz RBW. d) Sensitivity spectrum around the mechanical resonance. The black dashed line shows the sensitivity calculated from the thermomechanical noise model using the relevant parameter derived both experimentally and from the FEM  model}.
    \label{fig:Sensitivity}
\end{figure}

We assess the performance of the fabricated magnetometer using the experimental configuration 
%used to assess the performance of the magnetometer is 
illustrated in Figure~\ref{fig:Sensitivity}a). Here, light from a tunable laser (EXFO T100S-HP, 1500-1600\,nm) is coupled from a single-sided tapered optical fiber into the device's inverted tapered waveguide, in a similar manner to that presented in \cite{tiecke2015efficient, McQueen2025}. The reflected light is then directed to a photodetector (New Focus 1811) with the aid of a circulator (the detected optical power was approximately 1\,\textmu W). Monitoring the photodiode output while sweeping the laser wavelength reveals the optical resonance of the slot cavity, shown in Fig.~\ref{fig:Sensitivity}b). Fitting a Lorentzian lineshape to the reflected spectrum % ($R(\omega)$) around the cavity resonance ($\omega_0$) in order to extract the cavity energy loss rate ($\kappa$) in the form of:
% \begin{equation}
%         R(\omega) = R_\infty - \frac{(R_\infty - R_d) \left(\frac{\kappa}{2}\right)^2}{(\omega - \omega_0)^2 + \left(\frac{\kappa}{2}\right)^2}.
% \end{equation}
% Here $R_d$ ($R_\infty$) is the reflectivity at (far from) the the cavity resonance frequency. 
gives the cavity energy loss rate $\kappa/2\pi = 205$\,GHz and an optical quality factor ($Q_o = \omega_o/\kappa$) of $\sim10^3$. These are comparable to previously reported slotted photonic crystal cavities in SOI \cite{xu2013slotted, leijssen2015strong, wang2015single}.% indicating that the optical losses within the slot cavity are dominated by scattering at the photonic crystal mirror sites, rather than by the slot geometry.

In order to determine the magnetic field sensitivity, the laser wavelength was tuned to the side of the optical resonance and the photodiode output was monitored by network and spectrum analysers (Liquid Instruments Moku:Pro). In this configuration, changes to the optical resonance frequency produce amplitude modulations of the reflected light. Gas damping reduces the mechanical quality factor of the galfenol coated cantilever, and hence the overall sensitivity of our magnetometer. To improve performance, the device and the driving coil were placed in a vacuum chamber and sensitivity measurements were performed at a pressure of $1\times10^{-3}$\,mbar (see Supp. Sec.~S.4 for comparison to atmospheric pressure). This is a relatively modest vacuum, compatible with chip-scale vacuum packaging capabilities~ \cite{Sparks2003}. In the absence of a magnetic driving field, a broad mechanical resonance was observed, with a quality factor of $Q=635$. The peak frequency agreed with that of the eigenmode found using the FEM analysis discussed in Sec.~\ref{sec:design}. 

An AC magnetic field was generated by driving a home-made copper coil with a 10\,V\textsubscript{pp} sinusoidal signal, delivering a 410\,nT RMS AC field at the device. %The magnetically driven response with the field applied at the mechanical eigenfrequency delivered a signal that was approximately 40\,dB stronger than the thermomechanical noise at the same frequency (with a resolution bandwidth of 30\,Hz).
The sensitivity at this frequency ($\omega_\text{ref}$) is determined from the measured signal-to-noise ratio ($\text{SNR} = S(\omega_\text{ref})/N(\omega_\text{ref})$) of the response to this drive by \cite{Forstner2014}:
\begin{equation}
    B_{\text{min}}(\omega_{\text{ref}}) = \frac{B_{\text{AC}}}{\sqrt{\text{SNR} \times \text{RBW}}},
    \label{Eq-Bmin-Sa}
\end{equation}
where RBW is the spectrum analyzer resolution bandwidth and $B_{\text{AC}}$ is the applied magnetic field. Eq.~\eqref{Eq-Bmin-Sa} can be used in conjunction with the network response of the system to determine the sensitivity spectrum of the device ($B_{\text{min}}(\omega)$) by \cite{Forstner2014}: 

\begin{equation}
    B_{\text{min}}(\omega)= \sqrt{\frac{S(\omega_\text{ref})}{S(\omega)}\frac{N(\omega)}{N(\omega_\text{ref})}}\times B_{\text{min}}(\omega_{\text{ref}}),
\end{equation}
where $S(\omega)$ is the signal measured by the network analyzer and $N(\omega)$ is the noise floor measured with the frequency spectrum analyzer \cite{Forstner2012}. %\nb{Actually the spectrum analyzer noise floor and not the network ananlyzer noise floor}. 
The resulting sensitivity spectrum is shown in Fig.~\ref{fig:Sensitivity}d). The high optomechanical coupling suppressed the optical noise well below the thermomechanical noise due to %Thus, the magnetic sensitivity is determined by this dominant source, \textit{i.e.}, %caused by their combined thermal energy.
the random thermally-driven motion of the slot and cantilever, with the modelled thermomechanical noise limit (see Sec.~S3.3), agreeing well with measurements (dashed black curve, Fig.~\ref{fig:Sensitivity}d).\\ 
A peak sensitivity of 800\,p\TrtHz\ is observed at a frequency of 243\,kHz, corresponding to the mechanical resonance of the cantilever. This sensitivity is an improvement of several orders of magnitude when compared to previous on-chip waveguide-integrated optical magnetometers \cite{gotardo2023waveguide, wang2024giant, hoese2021integrated}.

%In this frequency band 

%A model of these noise contributions (see Sec.~S.4.3) is shown in Fig.~\ref{fig:Sensitivity}d); the calculated sensitivity spectrum agrees well with the experimentally determined values. 
% \textcolor{violet}{This modeling reveals several parameters which can be improved for increased sensitivity.
% Increasing the piezomagnetic coefficients (\textit{$\alpha_{33}$, \&, $d_{33}$}) would proportionally improve the sensitivity of the magnetometer. As discussed in Sec.~\ref{Sec:Fab}, galfenol thin films have reported up to a four times increase in these parameters. Recent reports of other magnetostrictive materials such as galfenol-B (FeGaB) thin films have demonstrated even higher values of $d_{33}$, as high as 150 nm/A \cite{dong2018characterization}. Employing such materials could improve the sensitivity by a factor of almost 40 to around 20 p\TrtHz.}

\section{Discussion}

The transition to a silicon-on-insulator (SOI) platform reported here represents a key enabler for chip-scale optomechanical magnetometry. Beyond its compatibility with photonic and electronic integration \cite{stojanovic2018monolithic}, SOI supports high-yield fabrication of suspended mechanical structures with submicron precision using standard lithographic tools \cite{zhang2023optomechanical}. This opens a viable path to monolithically integrated systems that combine optical readout, magnetic transduction, and CMOS-compatible control circuitry -- a level of integration that is not achievable with previous approaches. Previous optomechanical magnetometers have achieved peak sensitivity only within narrow frequency bands near mechanical resonances, with optical noise dominating across the rest of the spectrum. This significantly constrains possible applications, as most real-world use cases -- including navigation, magnetoencephalography, and geophysical surveying -- require operation at frequencies well below the mechanical resonance \cite{li2021cavity}.

By greatly increasing the optomechanical coupling strength, and thereby suppressing optical noise, our devices represent a significant step toward resolving this challenge. Although our sensors also exhibit peak sensitivity near resonance, the limiting factor is now thermomechanical noise rather than optical readout. Optical noise is typically dominated by shot noise near resonance and exhibits hard-to-suppress 1/f-like behaviour at low frequencies \cite{nelson20241/f}. By comparison, thermomechanical noise can be more readily mitigated through structural optimisation \cite{tsaturyan2017ultracoherent}.
In our devices, the dominant near-resonance noise arises from thermomechanical fluctuations of the galfenol-coated cantilever. This can be reduced by increasing the mechanical quality factor and by improving the transduction of magnetostrictive strain into transverse motion. The mechanical quality factor of the cantilever is not optimised, and is many orders-of-magnitude below the state-of-the-art -- now approaching
%with a value of 635, which is low compared to literature values that routinely exceed $10^6$ and can even approach 
$10^8$ at room temperature \cite{tsaturyan2017ultracoherent}. It is constrained primarily by clamping losses, which can be greatly suppressed using soft clamping techniques that are enabled by the transition to SOI \cite{sementilli2022nanomechanical}.

Away from resonance, the thermomechanical noise is dominated by motion of the slotted photonic crystal cavity. This is due to the presence of confined mechanical resonances that coexist with the desired confined optical mode. These high frequency resonances are strongly coupled to the optical mode. While the noise from the galfenol-coated cantilever has the same Lorentzian frequency dependence as the response to magnetic signals, and would result in a flat broadband magnetic sensitivity if fully dominant \cite{Yu2018}, the photonic crystal noise is effectively white across our full detection band. This results in degraded sensitivity away from the cantilever resonance frequency. 
Photonic crystal thermomechanical noise could be greatly reduced by designing the crystal to not support confined mechanical resonances \cite{enzian2023phononically}, or to suppress their motional amplitudes \cite{kraft2023suppressing}. Its contribution to the measured signal could be further reduced by increasing the mechanical coupling between the cantilever and the slot structure.

%This modeling reveals several parameters which can be improved for increased sensitivity.
Further, increasing the piezomagnetic coefficients (\textit{$\alpha_{33}$ \& $d_{33}$}) would proportionally improve the sensitivity of the magnetometer across the full frequency band. As discussed in Sec.~\ref{Sec:Fab}, improved galfenol thin films could allow a factor of four increase in these parameters \cite{García-Arribas2021}. Recent reports of other magnetostrictive materials such as galfenol-B (FeGaB) thin films have demonstrated even higher values of $d_{33}$, as high as 150\,nm/A \cite{dong2018characterization}. Employing such materials could improve the sensitivity by a factor over 45 to around 17\,p\TrtHz. Together, these strategies provide a feasible route to both improve the sensitivity to below 100\,f\TrtHz\ and extend high-sensitivity operation beyond narrow resonant bands, in a microscale, integrated magnetometer with microwatt optical power requirements. Further improved sensitivity could be achieved by increasing the device footprint to allow for a larger magnetostrictive element.

Our devices are already substantially more sensitive than previous waveguide-integrated optomechanical magnetometers, but are several orders of magnitude less sensitive than the best non-waveguide-integrated devices. We expect this sensitivity deficit to be overcome via the engineering strategies and material optimization discussed above. %, with further improvements possible by substituting galfenol with higher-performance magnetostrictive materials such as FeGaB \cite{hu2024picotesla}. 
Ultimately, optomechanical magnetometers may provide a compelling alternative to SQUIDs, OPMs, and fluxgate sensors -- offering comparable performance with dramatically lower size, weight, power, and cost.

\section{Methods} \label{sec:Methods}
\subsection{Fabrication details}
A silicon-on-insulator wafer (750\,\textmu m Si/3\,\textmu m $\text{SiO}_2$/0.22\,\textmu m Si) was cleaned and spincoated with a 350\,nm layer of electron-beam resist (ARP 6200.09). The device structure was defined using electron beam lithography (Raith EBPG 5150, 260\,\textmu C/cm\textsuperscript{2} dose) and developed with AR 600-546 (All Resist) then rinsed with o-Xylene and isopropyl alcohol (IPA). The exposed silicon was etched by reactive ion-etching (Oxford Instruments PlasmaPro NGP80, 50\,sccm $\text{SF}_6$ \& 100\,sccm $\text{CHF}_3$ at 82\,W RF power) forming the optical structures and cantilever in the Si device layer atop the buried oxide (BOX) silica layer. The devices were then cleaned before being released with hydroflouric acid (HF) vapor phase etching (VPE) at 35\textdegree C for 15~minutes. To enable complete release of the large triangular cantilevers (200\,\textmu m base $\times$ 100\,\textmu m height), holes were added to allow the HF vapor to reach the silica underneath.

Photoresist (AZ nLoF 2020) was diluted in equal parts with edge bead removal (EBR) solvent to significantly reduce its viscosity. This diluted photoresist solution was spin-coated onto the released devices at a speed of 1000\,RPM, removing the solution from the top of the device layer while filling the volume of the removed BOX. This was followed by baking at 100\textdegree C for 60\,s to partially solidify the solution. The procedure was then repeated to ensure complete substitution of the removed BOX and provide a supporting layer underneath the released devices. The galfenol deposition area was then defined by patterning with a typical photolithography method by spin-coating undiluted AZ~ nLoF~2020, photolithographic exposure (MLA150, Heidelberg Instruments Maskless aligner), and development in AZ~726. The magnetostrictive thin-film was then deposited by DC sputtering (Temescal FC-2000) a 5\,nm tantalum adhesion layer, 16\,nm copper seeding layer, 600\,nm galfenol layer, and a 5\,nm tantalum capping layer to protect against oxidation. Excess galfenol was removed in acetone via a magnetically assisted liftoff process. The fabrication process was completed by transferring the chip with freestanding devices into IPA followed by carbon dioxide critical point drying (Leica EM 300CPD) to eliminate stiction.

\backmatter

\bmhead{Supplementary information}

See Supplemental Material for supporting content, including additional information on fabrication procedures, investigations of the effects of HF on galfenol, fiber bonding to the integrated waveguide, further modeling of the device and the thermomechanical limited sensitivity, and details of mechanical modes with varying pressure. 

\bmhead{Acknowledgements}

The authors acknowledge the facilities, and the scientific and technical assistance, of the Australian Microscopy \& Microanalysis Research Facility at the Centre for Microscopy and Microanalysis, The University of Queensland. This work was performed in part at the Queensland node of the Australian National Fabrication Facility, a company established under the National Collaborative Research Infrastructure Strategy to provide nano- and microfabrication facilities for Australia's researchers. The Commonwealth of Australia (represented by the Defence Science and Technology Group) supports this research through a Defence Science Partnerships agreement. This work was funded under the NGTF and is being delivered through the ASCA. This work was financially supported by the Australian Research Council (ARC) Centres of excellence for Engineered Quantum systems (EQUS, Grant No. CE170100009) and Quantum Biotechnology (Grant No. CE230100021). We thank the Queensland Defence Science Alliance (QDSA) for financial support of the project through the 2023 QDSA Collaborative Research Grant (CRG) funding round. The authors acknowledge the highly valuable advice and support provided by Rodney Appleby and financial support by Orica Australia \textit{Pty Ltd}. This project was funded by the Queensland Government through the Department of Environment, Tourism, Science and Innovation’s (DETSI) Quantum 2032 Challenge Program. The program accelerates quantum sportstech, connects Queensland's research sector with industry, and showcases the state's quantum capabilities as part of Brisbane 2032's legacy.

\section*{Declarations}
The authors declare no conflicts of interest.


\bibliography{references}% common bib file

@article{abramovici1992ligo,
  title={{LIGO}: The laser interferometer gravitational-wave observatory},
  author={Abramovici, Alex and Althouse, William E and Drever, Ronald WP and G{\"u}rsel, Yekta and Kawamura, Seiji and Raab, Frederick J and Shoemaker, David and Sievers, Lisa and Spero, Robert E and Thorne, Kip S and others},
  journal={Science},
  volume={256},
  number={5055},
  pages={325--333},
  year={1992},
  publisher={American Association for the Advancement of Science}
}

@article{barzanjeh2022optomechanics,
  title={Optomechanics for quantum technologies},
  author={Barzanjeh, Shabir and Xuereb, Andr{\'e} and Gr{\"o}blacher, Simon and Paternostro, Mauro and Regal, Cindy A and Weig, Eva M},
  journal={Nature Physics},
  volume={18},
  number={1},
  pages={15--24},
  year={2022},
  publisher={Nature Publishing Group UK London}
}

@article{brady2023entanglement,
  title={Entanglement-enhanced optomechanical sensor array with application to dark matter searches},
  author={Brady, Anthony J and Chen, Xin and Xia, Yi and Manley, Jack and Dey Chowdhury, Mitul and Xiao, Kewen and Liu, Zhen and Harnik, Roni and Wilson, Dalziel J and Zhang, Zheshen and others},
  journal={Communications Physics},
  volume={6},
  number={1},
  pages={237},
  year={2023},
  publisher={Nature Publishing Group UK London}
}

@article{baker2024optomechanical,
  title={Optomechanical dark matter instrument for direct detection},
  author={Baker, Christopher G and Bowen, Warwick P and Cox, Peter and Dolan, Matthew J and Goryachev, Maxim and Harris, Glen},
  journal={Physical Review D},
  volume={110},
  number={4},
  pages={043005},
  year={2024},
  publisher={APS}
}

@article{gavartin2012hybrid,
  title={A hybrid on-chip optomechanical transducer for ultrasensitive force measurements},
  author={Gavartin, Emanuel and Verlot, Pierre and Kippenberg, Tobias J},
  journal={Nature Nanotechnology},
  volume={7},
  number={8},
  pages={509--514},
  year={2012},
  publisher={Nature Publishing Group UK London}
}

@article{mason2019continuous,
  title={Continuous force and displacement measurement below the standard quantum limit},
  author={Mason, David and Chen, Junxin and Rossi, Massimiliano and Tsaturyan, Yeghishe and Schliesser, Albert},
  journal={Nature Physics},
  volume={15},
  number={8},
  pages={745--749},
  year={2019},
  publisher={Nature Publishing Group UK London}
}

@article{schliesser2008high,
  title={High-sensitivity monitoring of micromechanical vibration using optical whispering gallery mode resonators},
  author={Schliesser, Albert and Anetsberger, Georg and Rivi{\`e}re, R{\'e}mi and Arcizet, Olivier and Kippenberg, Tobias J},
  journal={New Journal of Physics},
  volume={10},
  number={9},
  pages={095015},
  year={2008},
  publisher={IOP Publishing}
}

@article{cao2025integrated,
  title={Integrated optomechanical ultrasonic sensors with nano-Pascal-level sensitivity},
  author={Cao, Xuening and Yang, Hao and Wang, Min and Hu, Zhi-Gang and Wu, Zu-Lei and Wang, Yuanlei and Liu, Jian-Fei and Zhou, Xin and Li, Jincheng and Lao, Chenghao and others},
  journal={arXiv preprint arXiv:2506.20219},
  year={2025}
}

@article{bawden2025precision,
  title={Precision optomechanical accelerometer via hybrid test mass integration},
  author={Bawden, Nathaniel and Carey, Benjamin J and Yeo, Poh-Meng and Arora, Nishta and Sementilli, Leo and Valenzuela, Victor M and Romero, Erick and Harris, Glen I and Wegener, Margaret and Bowen, Warwick P},
  journal={arXiv preprint arXiv:2508.16088},
  year={2025}
}

@article{wasserman2022cryogenic,
  title={Cryogenic and hermetically sealed packaging of photonic chips for optomechanics},
  author={Wasserman, WW and Harrison, RA and Harris, GI and Sawadsky, Andreas and Sfendla, YL and Bowen, WP and Baker, CG},
  journal={Optics Express},
  volume={30},
  number={17},
  pages={30822--30831},
  year={2022},
  publisher={Optica Publishing Group}
}

@article{almeida2004guiding,
  title={Guiding and confining light in void nanostructure},
  author={Almeida, Vilson R and Xu, Qianfan and Barrios, Carlos A and Lipson, Michal},
  journal={Optics Letters},
  volume={29},
  number={11},
  pages={1209--1211},
  year={2004},
  publisher={Optical Society of America}
}

@article{aspelmeyer2014cavity,
  title={Cavity optomechanics},
  author={Aspelmeyer, Markus and Kippenberg, Tobias J and Marquardt, Florian},
  journal={Reviews of Modern Physics},
  volume={86},
  number={4},
  pages={1391--1452},
  year={2014},
  publisher={APS}
}

@article{sun2015single,
  title={Single-chip microprocessor that communicates directly using light},
  author={Sun, Chen and Wade, Mark T and Lee, Yunsup and Orcutt, Jason S and Alloatti, Luca and Georgas, Michael S and Waterman, Andrew S and Shainline, Jeffrey M and Avizienis, Rimas R and Lin, Sen and others},
  journal={Nature},
  volume={528},
  number={7583},
  pages={534--538},
  year={2015},
  publisher={Nature Publishing Group UK London}
}

@article{lee1997dry,
  title={Dry release for surface micromachining with {HF} vapor-phase etching},
  author={Lee, Y-I and Park, K-H and Lee, Jonghyun and Lee, C-S and Yoo, Hyung Joun and Kim, C-J and Yoon, Y-S},
  journal={Journal of Microelectromechanical Systems},
  volume={6},
  number={3},
  pages={226--233},
  year={1997},
  publisher={IEEE}
}

@article{bekker2018free,
  title={Free spectral range electrical tuning of a high quality on-chip microcavity},
  author={Bekker, Christiaan and Baker, Christopher G and Kalra, Rachpon and Cheng, Han-Hao and Li, Bei-Bei and Prakash, Varun and Bowen, Warwick P},
  journal={Optics Express},
  volume={26},
  number={26},
  pages={33649--33670},
  year={2018},
  publisher={Optical Society of America}
}

@article{su2023scalability,
  title={Scalability of large-scale photonic integrated circuits},
  author={Su, Yikai and He, Yu and Guo, Xuhan and Xie, Weiqiang and Ji, Xingchen and Wang, Hongwei and Cai, Xinlun and Tong, Limin and Yu, Siyuan},
  journal={ACS Photonics},
  volume={10},
  number={7},
  pages={2020--2030},
  year={2023},
  publisher={ACS Publications}
}

@article{Forstner2014,
author = {Forstner, Stefan and Sheridan, Eoin and Knittel, Joachim and Humphreys, Christopher L. and Brawley, George A. and Rubinsztein-Dunlop, Halina and Bowen, Warwick P.},
title = {Ultrasensitive Optomechanical Magnetometry},
journal = {Advanced Materials},
volume = {26},
number = {36},
pages = {6348-6353},
year = {2014}
}

@article{Savukov2013,
    author = {Savukov, I. and Karaulanov, T.},
    title = "{Magnetic-resonance imaging of the human brain with an atomic magnetometer}",
    journal = {Applied Physics Letters},
    volume = {103},
    number = {4},
    year = {2013},
    month = {07},
    issn = {0003-6951}
}

@Article{Bennett2021,
AUTHOR = {Bennett, James S. and Vyhnalek, Brian E. and Greenall, Hamish and Bridge, Elizabeth M. and Gotardo, Fernando and Forstner, Stefan and Harris, Glen I. and Miranda, Félix A. and Bowen, Warwick P.},
TITLE = {Precision Magnetometers for Aerospace Applications: A Review},
JOURNAL = {Sensors},
VOLUME = {21},
YEAR = {2021},
NUMBER = {16},
ARTICLE-NUMBER = {5568},
PubMedID = {34451010},
ISSN = {1424-8220}
}

@article{Li:20,
author = {Bei-Bei Li and George Brawley and Hamish Greenall and Stefan Forstner and Eoin Sheridan and Halina Rubinsztein-Dunlop and Warwick P. Bowen},
journal = {Photonics Research},
number = {7},
pages = {1064--1071},
publisher = {Optica Publishing Group},
title = {Ultrabroadband and sensitive cavity optomechanical magnetometry},
volume = {8},
month = {Jul},
year = {2020},
}

@article{Forstner2012,
  title = {Cavity Optomechanical Magnetometer},
  author = {Forstner, S. and Prams, S. and Knittel, J. and van Ooijen, E. D. and Swaim, J. D. and Harris, G. I. and Szorkovszky, A. and Bowen, W. P. and Rubinsztein-Dunlop, H.},
  journal = {Physical Review Letters},
  volume = {108},
  issue = {12},
  pages = {120801},
  numpages = {5},
  year = {2012},
  month = {Mar},
  publisher = {American Physical Society},
  url = {https://link.aps.org/doi/10.1103/PhysRevLett.108.120801}
}

@article{Li:18Quant,
author = {Bei-Bei Li and Jan B\'{i}lek and Ulrich B. Hoff and Lars S. Madsen and Stefan Forstner and Varun Prakash and Clemens Sch\"{a}fermeier and Tobias Gehring and Warwick P. Bowen and Ulrik L. Andersen},
journal = {Optica},
number = {7},
pages = {850--856},
publisher = {Optica Publishing Group},
title = {Quantum enhanced optomechanical magnetometry},
volume = {5},
month = {Jul},
year = {2018},
url = {https://opg.optica.org/optica/abstract.cfm?URI=optica-5-7-850},
}

@Article{Yu2018,
AUTHOR = {Yu, Yimin and Forstner, Stefan and Rubinsztein-Dunlop, Halina and Bowen, Warwick Paul},
TITLE = {Modelling of Cavity Optomechanical Magnetometers},
JOURNAL = {Sensors},
VOLUME = {18},
YEAR = {2018},
NUMBER = {5},
ARTICLE-NUMBER = {1558},
PubMedID = {29758002},
ISSN = {1424-8220}
}

@article{Nivedita2017galf,
title = {Enhancement of magnetostrictive properties of Galfenol thin films},
journal = {Journal of Magnetism and Magnetic Materials},
volume = {451},
pages = {300-304},
year = {2018},
issn = {0304-8853},
author = {Lalitha Raveendran Nivedita and Palanisamy Manivel and Ramanathaswamy Pandian and S. Murugesan and Nicola Ann Morley and K. Asokan and Ramasamy Thangavelu {Rajendra Kumar}}
}

@article{Purdy2017,
    author = {Purdy, T. P. and Grutter, K. E. and Srinivasan, K. and Taylor, J. M.},
    title = {Quantum correlations from a room-temperature optomechanical cavity},
    journal = {Science},
    volume = {356},
    number = {6344},
    year = {2017},
    pages = {1265-1268}
}

@article{adolphi2010improvement,
  title={Improvement of sputtered Galfenol thin films for sensor applications},
  author={Adolphi, B and McCord, J and Bertram, M and Oertel, CG and Merkel, U and Marschner, U and Sch{\"a}fer, R and Wenzel, C and Fischer, WJ},
  journal={Smart Materials and Structures},
  volume={19},
  number={5},
  pages={055013},
  year={2010},
  publisher={IOP Publishing}
}

@article{wang2024giant,
  title={Giant magneto-photoluminescence at ultralow field in organic microcrystal arrays for on-chip optical magnetometer},
  author={Wang, Hong and Yin, Baipeng and Bai, Junli and Wei, Xiao and Huang, Wenjin and Chang, Qingda and Jia, Hao and Chen, Rui and Zhai, Yaxin and Wu, Yuchen and others},
  journal={Nature Communications},
  volume={15},
  number={1},
  pages={3995},
  year={2024},
  publisher={Nature Publishing Group UK London}
}

@article{gotardo2023waveguide,
  title={Waveguide-integrated chip-scale optomechanical magnetometer},
  author={Gotardo, Fernando and Carey, Benjamin J and Greenall, Hamish and Harris, Glen I and Romero, Erick and Bulla, Douglas and Bridge, Elizabeth M and Bennett, James S and Foster, Scott and Bowen, Warwick P},
  journal={Optics Express},
  volume={31},
  number={23},
  pages={37663--37672},
  year={2023},
  publisher={Optica Publishing Group}
}

@article{greenall2024QPM,
  title={Quantitative profilometric measurement of magnetostriction in thin-films},
  author={Greenall, Hamish and Carey, Benjamin J and Bulla, Douglas and Gotardo, Fernando and Harris, Glen I and Bennett, James S and Foster, Scott and Bowen, Warwick P},
  journal={Applied Surface Science},
  volume={662},
  pages={160105},
  year={2024},
  publisher={Elsevier}
}

@article{hu2024picotesla,
  title={Picotesla-sensitivity microcavity optomechanical magnetometry},
  author={Hu, Zhi-Gang and Gao, Yi-Meng and Liu, Jian-Fei and Yang, Hao and Wang, Min and Lei, Yuechen and Zhou, Xin and Li, Jincheng and Cao, Xuening and Liang, Jinjing and others},
  journal={Light: Science \& Applications},
  volume={13},
  number={1},
  pages={279},
  year={2024},
  publisher={Nature Publishing Group UK London}
}

@article{brookes2022magnetoencephalography,
  title={Magnetoencephalography with optically pumped magnetometers ({OPM-MEG}): the next generation of functional neuroimaging},
  author={Brookes, Matthew J and Leggett, James and Rea, Molly and Hill, Ryan M and Holmes, Niall and Boto, Elena and Bowtell, Richard},
  journal={Trends in Neurosciences},
  volume={45},
  number={8},
  pages={621--634},
  year={2022},
  publisher={Elsevier}
}

@article{murzin2020ultrasensitive,
  title={Ultrasensitive magnetic field sensors for biomedical applications},
  author={Murzin, Dmitry and Mapps, Desmond J and Levada, Kateryna and Belyaev, Victor and Omelyanchik, Alexander and Panina, Larissa and Rodionova, Valeria},
  journal={Sensors},
  volume={20},
  number={6},
  pages={1569},
  year={2020},
  publisher={MDPI}
}

@article{stolz2022squids,
  title={SQUIDs for magnetic and electromagnetic methods in mineral exploration},
  author={Stolz, Ronny and Schiffler, Markus and Becken, Michael and Thiede, Anneke and Schneider, Michael and Chubak, Glenn and Marsden, Paul and Bergshjorth, Ana Bra{\~n}a and Schaefer, Markus and Terblanche, Ockert},
  journal={Mineral Economics},
  volume={35},
  number={3},
  pages={467--494},
  year={2022},
  publisher={Springer}
}

@article{lu2023recent,
  title={Recent progress of atomic magnetometers for geomagnetic applications},
  author={Lu, Yuantian and Zhao, Tian and Zhu, Wanhua and Liu, Leisong and Zhuang, Xin and Fang, Guangyou and Zhang, Xiaojuan},
  journal={Sensors},
  volume={23},
  number={11},
  pages={5318},
  year={2023},
  publisher={MDPI}
}

@article{basiri2019precision,
  title={Precision ultrasound sensing on a chip},
  author={Basiri-Esfahani, Sahar and Armin, Ardalan and Forstner, Stefan and Bowen, Warwick P},
  journal={Nature Communications},
  volume={10},
  number={1},
  pages={132},
  year={2019},
  publisher={Nature Publishing Group UK London}
}

@article{navarathna2024silicon,
  title={Silicon double-disk optomechanical resonators from wafer-scale double-layered silicon-on-insulator},
  author={Navarathna, Amy and Carey, Benjamin J and Bennett, James S and Khademi, Soroush and Bowen, Warwick P},
  journal={Optics Express},
  volume={32},
  number={23},
  pages={41376--41389},
  year={2024},
  publisher={Optica Publishing Group}
}

@article{xu2013slotted,
  title={Slotted photonic crystal nanobeam cavity with parabolic modulated width stack for refractive index sensing},
  author={Xu, Peipeng and Yao, Kaiyuan and Zheng, Jiajiu and Guan, Xiaowei and Shi, Yaocheng},
  journal={Optics Express},
  volume={21},
  number={22},
  pages={26908--26913},
  year={2013},
  publisher={Optica Publishing Group}
}

@article{leijssen2015strong,
  title={Strong optomechanical interactions in a sliced photonic crystal nanobeam},
  author={Leijssen, Rick and Verhagen, Ewold},
  journal={Scientific Reports},
  volume={5},
  number={1},
  pages={15974},
  year={2015},
  publisher={Nature Publishing Group UK London}
}

@article{wang2015single,
  title={Single-nanoparticle detection with slot-mode photonic crystal cavities},
  author={Wang, Cheng and Quan, Qimin and Kita, Shota and Li, Yihang and Lon{\v{c}}ar, Marko},
  journal={Applied Physics Letters},
  volume={106},
  number={26},
  year={2015},
  publisher={AIP Publishing}
}

@article{connerney2015maven,
  title={The {MAVEN} magnetic field investigation},
  author={Connerney, JEP and Espley, J and Lawton, P and Murphy, S and Odom, J and Oliversen, R and Sheppard, Dave},
  journal={Space Science Reviews},
  volume={195},
  number={1},
  pages={257--291},
  year={2015},
  publisher={Springer}
}

@article{storm2017ultra,
  title={An ultra-sensitive and wideband magnetometer based on a superconducting quantum interference device},
  author={Storm, Jan-Hendrik and H{\"o}mmen, Peter and Drung, Dietmar and K{\"o}rber, Rainer},
  journal={Applied Physics Letters},
  volume={110},
  number={7},
  year={2017},
  publisher={AIP Publishing}
}

@article{herrera2016recent,
  title={Recent advances of {MEMS} resonators for {Lorentz} force based magnetic field sensors: design, applications and challenges},
  author={Herrera-May, Agust{\'\i}n Leobardo and Soler-Balcazar, Juan Carlos and V{\'a}zquez-Leal, H{\'e}ctor and Mart{\'\i}nez-Castillo, Jaime and Vigueras-Zu{\~n}iga, Marco Osvaldo and Aguilera-Cort{\'e}s, Luz Antonio},
  journal={Sensors},
  volume={16},
  number={9},
  pages={1359},
  year={2016},
  publisher={MDPI}
}

@article{borna2020non,
  title={Non-invasive functional-brain-imaging with an {OPM}-based magnetoencephalography system},
  author={Borna, Amir and Carter, Tony R and Colombo, Anthony P and Jau, Yuan-Yu and McKay, Jim and Weisend, Michael and Taulu, Samu and Stephen, Julia M and Schwindt, Peter DD},
  journal={Plos one},
  volume={15},
  number={1},
  pages={e0227684},
  year={2020},
  publisher={Public Library of Science San Francisco, CA USA}
}

@article{tiecke2015efficient,
  title={Efficient fiber-optical interface for nanophotonic devices},
  author={Tiecke, TG and Nayak, KP and Thompson, Jeffrey Douglas and Peyronel, T and de Leon, Nathalie P and Vuleti{\'c}, V and Lukin, MD},
  journal={Optica},
  volume={2},
  number={2},
  pages={70--75},
  year={2015},
  publisher={Optica Publishing Group}
}

@article{hoese2021integrated,
  title={Integrated magnetometry platform with stackable waveguide-assisted detection channels for sensing arrays},
  author={Hoese, Michael and Koch, Michael K and Bharadwaj, Vibhav and Lang, Johannes and Hadden, John P and Yoshizaki, Reina and Giakoumaki, Argyro N and Ramponi, Roberta and Jelezko, Fedor and Eaton, Shane M and others},
  journal={Physical Review Applied},
  volume={15},
  number={5},
  pages={054059},
  year={2021},
  publisher={APS}
}

@article{Winger_2011,
author = {M. Winger and T. D. Blasius and T. P. Mayer Alegre and A. H. Safavi-Naeini and S. Meenehan and J. Cohen and S. Stobbe and O. Painter},
journal = {Optics Express},
number = {25},
pages = {24905--24921},
publisher = {Optica Publishing Group},
title = {A chip-scale integrated cavity-electro-optomechanics platform},
volume = {19},
month = {Dec},
year = {2011},
}

@article{woolf2013optomechanical,
  title={Optomechanical and photothermal interactions in suspended photonic crystal membranes},
  author={Woolf, David and Hui, Pui-Chuen and Iwase, Eiji and Khan, Mughees and Rodriguez, Alejandro W and Deotare, Parag and Bulu, Irfan and Johnson, Steven G and Capasso, Federico and Loncar, Marko},
  journal={Optics Express},
  volume={21},
  number={6},
  pages={7258--7275},
  year={2013},
  publisher={Optica Publishing Group}
}

@inproceedings{zhang2023optomechanical,
  title={Optomechanical cavities in silicon-on-insulator},
  author={Zhang, Jianhao and Nu{\~n}o-Ruano, Paula and Le Roux, Xavier and Cassan, Eric and Marris-Morini, Delphine and Vivien, Laurent and Lanzillotti-Kimura, Norberto Daniel and Ramos, Carlos},
  booktitle={Integrated Optics: Devices, Materials, and Technologies XXVII},
  volume={12424},
  pages={12--16},
  year={2023},
  organization={SPIE}
}

@Article{Xia_Optomech_review_2020,
AUTHOR = {Xia, Ji and Qiao, Qifeng and Zhou, Guangcan and Chau, Fook Siong and Zhou, Guangya},
TITLE = {Opto-Mechanical Photonic Crystal Cavities for Sensing Application},
JOURNAL = {Applied Sciences},
VOLUME = {10},
YEAR = {2020},
NUMBER = {20},
ARTICLE-NUMBER = {7080},
ISSN = {2076-3417}
}

@article{McQueen2025,
  title={Fibre-coupled photonic crystal hydrophone},
  author={McQueen, Lauren R and Bawden, Nathaniel and Carey, Benjamin J and Marinkovi{\'c}, Igor and Bowen, Warwick P and Harris, Glen I},
  journal={Optics Express},
  volume={33},
  number={12},
  pages={25910--25921},
  year={2025},
  publisher={Optica Publishing Group}
}

@article{stele2023drone,
  title={Drone-based magnetometer prospection for archaeology},
  author={Stele, Andreas and Kaub, Leon and Linck, Roland and Schikorra, Markus and Fassbinder, J{\"o}rg WE},
  journal={Journal of Archaeological Science},
  volume={158},
  pages={105818},
  year={2023},
  publisher={Elsevier}
}

@article{velha2007ultra,
  title={Ultra-high {Q/V} {Fabry-Perot} microcavity on {SOI} substrate},
  author={Velha, Philippe and Picard, E and Charvolin, T and Hadji, E and Rodier, Jean-Claude and Lalanne, Philippe and Peyrade, David},
  journal={Optics Express},
  volume={15},
  number={24},
  pages={16090--16096},
  year={2007},
  publisher={Optica Publishing Group}
}

@article{xu2024subpicotesla,
  title={Subpicotesla Optomechanical Magnetometry},
  author={Xu, An-Ning and Li, Yifan and Li, Xiangliang and Liu, Bei and Liu, Yong-Chun},
  journal={Physical Review Letters},
  volume={133},
  number={15},
  pages={153601},
  year={2024},
  publisher={APS}
}

@ARTICLE{Sparks2003,
  author={Sparks, D.R. and Massoud-Ansari, S. and Najafi, N.},
  journal={IEEE Transactions on Advanced Packaging}, 
  title={Chip-level vacuum packaging of micromachines using NanoGetters}, 
  year={2003},
  volume={26},
  number={3},
  pages={277-282}
}

@article{li2021cavity,
  title={Cavity optomechanical sensing},
  author={Li, Bei-Bei and Ou, Lingfeng and Lei, Yuechen and Liu, Yong-Chun},
  journal={Nanophotonics},
  volume={10},
  number={11},
  pages={2799--2832},
  year={2021},
  publisher={De Gruyter}
}

@article{dong2018characterization,
  title={Characterization of magnetomechanical properties in {FeGaB} thin films},
  author={Dong, Cunzheng and Li, Menghui and Liang, Xianfeng and Chen, Huaihao and Zhou, Haomiao and Wang, Xinjun and Gao, Yuan and McConney, Michael E and Jones, John G and Brown, Gail J and others},
  journal={Applied Physics Letters},
  volume={113},
  number={26},
  year={2018},
  publisher={AIP Publishing}
}

@Inbook{García-Arribas2021,
author="Garc{\'i}a-Arribas, Alfredo",
editor="Franco, Victorino
and Dodrill, Brad",
title="Magnetostrictive Materials",
bookTitle="Magnetic Measurement Techniques for Materials Characterization",
year="2021",
publisher="Springer International Publishing",
address="Cham",
pages="727--750"
}

@article{luo2024magnetoelectric,
  title={Magnetoelectric microelectromechanical and nanoelectromechanical systems for the IoT},
  author={Luo, Bin and Will-Cole, AR and Dong, Cunzheng and He, Yifan and Liu, Xiaxin and Lin, Hwaider and Huang, Rui and Shi, Xiaoling and McConney, Michael and Page, Michael and others},
  journal={Nature Reviews Electrical Engineering},
  volume={1},
  number={5},
  pages={317--334},
  year={2024},
  publisher={Nature Publishing Group UK London}
}

@article{dai2021recent,
  title={Recent progress on the corrosion behavior of metallic materials in HF solution},
  author={Dai, Hailong and Shi, Shouwen and Yang, Lin and Guo, Can and Chen, Xu},
  journal={Corrosion Reviews},
  volume={39},
  number={4},
  pages={313--337},
  year={2021},
  publisher={De Gruyter}
}

@article{stojanovic2018monolithic,
  title={Monolithic silicon-photonic platforms in state-of-the-art {CMOS SOI} processes},
  author={Stojanovi{\'c}, Vladimir and Ram, Rajeev J and Popovi{\'c}, Milos and Lin, Sen and Moazeni, Sajjad and Wade, Mark and Sun, Chen and Alloatti, Luca and Atabaki, Amir and Pavanello, Fabio and others},
  journal={Optics Express},
  volume={26},
  number={10},
  pages={13106--13121},
  year={2018},
  publisher={Optical Society of America}
}

@article{atabaki2018integrating,
  title={Integrating photonics with silicon nanoelectronics for the next generation of systems on a chip},
  author={Atabaki, Amir H and Moazeni, Sajjad and Pavanello, Fabio and Gevorgyan, Hayk and Notaros, Jelena and Alloatti, Luca and Wade, Mark T and Sun, Chen and Kruger, Seth A and Meng, Huaiyu and others},
  journal={Nature},
  volume={556},
  number={7701},
  pages={349--354},
  year={2018},
  publisher={Nature Publishing Group UK London}
}

@article{savukov2005nmr,
  title={{NMR} detection with an atomic magnetometer},
  author={Savukov, IM and Romalis, Michael V},
  journal={Physical Review Letters},
  volume={94},
  number={12},
  pages={123001},
  year={2005},
  publisher={APS}
}

@article{shekhar2024roadmapping,
  title={Roadmapping the next generation of silicon photonics},
  author={Shekhar, Sudip and Bogaerts, Wim and Chrostowski, Lukas and Bowers, John E and Hochberg, Michael and Soref, Richard and Shastri, Bhavin J},
  journal={Nature Communications},
  volume={15},
  number={1},
  pages={751},
  year={2024},
  publisher={Nature Publishing Group UK London}
}

@article{liu2020design,
  title={Design and fabrication of low-deformation micro-bolometers for THz detectors},
  author={Liu, Ziji and Liang, Zhiqing and Tang, Wen and Xu, Xiangdong},
  journal={Infrared Physics \& Technology},
  volume={105},
  pages={103241},
  year={2020},
  publisher={Elsevier}
}

@article{dao2019mems,
  title={{MEMS}-based wavelength-selective bolometers},
  author={Dao, Thang Duy and Doan, Anh Tung and Ishii, Satoshi and Yokoyama, Takahiro and {\O}rjan, Handeg{\aa}rd Sele and Ngo, Dang Hai and Ohki, Tomoko and Ohi, Akihiko and Wada, Yoshiki and Niikura, Chisato and others},
  journal={Micromachines},
  volume={10},
  number={6},
  pages={416},
  year={2019},
  publisher={MDPI}
}

@inproceedings{ramasamy2011magneto,
  title={Magneto-mechanical MEMS sensors for bio-detection},
  author={Ramasamy, M and Liang, C and Prorok, BC},
  booktitle={MEMS and Nanotechnology, Volume 2: Proceedings of the 2010 Annual Conference on Experimental and Applied Mechanics},
  pages={9--15},
  year={2011},
  organization={Springer}
}

@article{dreher2012surface,
  title={Surface acoustic wave driven ferromagnetic resonance in nickel thin films: {Theory} and experiment},
  author={Dreher, Lukas and Weiler, Mathias and Pernpeintner, Matthias and Huebl, Hans and Gross, Rudolf and Brandt, Martin S and Goennenwein, Sebastian TB},
  journal={Physical Review B},
  volume={86},
  number={13},
  pages={134415},
  year={2012},
  publisher={APS}
}

@article{tereshchenko2025direct,
  title={Direct Spin-Imaging Detector Based on Freestanding Magnetic Nanomembranes},
  author={Tereshchenko, OE and Bakin, VV and Stepanov, SA and Golyashov, VA and Mikaeva, AS and Kustov, DA and Rusetsky, VS and Rozhkov, SA and Scheibler, HE and Demin, A Yu},
  journal={Physical Review Letters},
  volume={134},
  number={15},
  pages={157002},
  year={2025},
  publisher={APS}
}

@article{tsaturyan2017ultracoherent,
  title={Ultracoherent nanomechanical resonators via soft clamping and dissipation dilution},
  author={Tsaturyan, Yeghishe and Barg, Andreas and Polzik, Eugene S and Schliesser, Albert},
  journal={Nature Nanotechnology},
  volume={12},
  number={8},
  pages={776--783},
  year={2017},
  publisher={Nature Publishing Group UK London}
}

@article{o2010quantum,
  title={Quantum ground state and single-phonon control of a mechanical resonator},
  author={O’Connell, Aaron D and Hofheinz, Max and Ansmann, Markus and Bialczak, Radoslaw C and Lenander, Mike and Lucero, Erik and Neeley, Matthew and Sank, Daniel and Wang, H and Weides, Martin and others},
  journal={Nature},
  volume={464},
  number={7289},
  pages={697--703},
  year={2010},
  publisher={Nature Publishing Group UK London}
}

@article{pirkkalainen2013hybrid,
  title={Hybrid circuit cavity quantum electrodynamics with a micromechanical resonator},
  author={Pirkkalainen, J-M and Cho, SU and Li, Jian and Paraoanu, GS and Hakonen, PJ and Sillanp{\"a}{\"a}, MA},
  journal={Nature},
  volume={494},
  number={7436},
  pages={211--215},
  year={2013},
  publisher={Nature Publishing Group UK London}
}

@article{shi2019study,
  title={A study of high piezomagnetic {(Fe-Ga/Fe-Ni)} multilayers for magnetoelectric device},
  author={Shi, Jiaxing and Wu, Ming and Hu, Wenlong and Lu, Cifu and Mu, Xing and Zhu, Jie},
  journal={Journal of Alloys and Compounds},
  volume={806},
  pages={1465--1468},
  year={2019},
  publisher={Elsevier}
}

@article{nelson20241/f,
  title={1/f Noise Mitigation in an Opto-Mechanical Sensor with a {Fabry--P{\'e}rot} Interferometer},
  author={Nelson, Andrea M and Sanjuan, Jose and Guzm{\'a}n, Felipe},
  journal={Sensors},
  volume={24},
  number={6},
  pages={1969},
  year={2024},
  publisher={MDPI}
}

@article{enzian2023phononically,
  title={Phononically shielded photonic-crystal mirror membranes for cavity quantum optomechanics},
  author={Enzian, Georg and Wang, Zihua and Simonsen, Anders and Mathiassen, Jonas and Vibel, Toke and Tsaturyan, Yeghishe and Tagantsev, Alexander and Schliesser, Albert and Polzik, Eugene S},
  journal={Optics Express},
  volume={31},
  number={8},
  pages={13040--13052},
  year={2023},
  publisher={Optica Publishing Group}
}

@article{kraft2023suppressing,
  title={Suppressing the mechanochromism of flexible photonic crystals},
  author={Kraft, Fabio A and Harwardt, Katharina and Schardt, Jan and Nowotka, Dirk and Gerken, Martina},
  journal={Optics Express},
  volume={31},
  number={4},
  pages={6281--6295},
  year={2023},
  publisher={Optica Publishing Group}
}

@article{sementilli2022nanomechanical,
  title={Nanomechanical dissipation and strain engineering},
  author={Sementilli, Leo and Romero, Erick and Bowen, Warwick P},
  journal={Advanced Functional Materials},
  volume={32},
  number={3},
  pages={2105247},
  year={2022},
  publisher={Wiley Online Library}
}

@article{devience2015nanoscale,
  title={Nanoscale {NMR} spectroscopy and imaging of multiple nuclear species},
  author={DeVience, Stephen J and Pham, Linh M and Lovchinsky, Igor and Sushkov, Alexander O and Bar-Gill, Nir and Belthangady, Chinmay and Casola, Francesco and Corbett, Madeleine and Zhang, Huiliang and Lukin, Mikhail and others},
  journal={Nature Nanotechnology},
  volume={10},
  number={2},
  pages={129--134},
  year={2015},
  publisher={Nature Publishing Group UK London}
}

@article{armani2003ultra,
  title={Ultra-high-{Q} toroid microcavity on a chip},
  author={Armani, DK and Kippenberg, TJ and Spillane, SM and Vahala, KJ},
  journal={Nature},
  volume={421},
  number={6926},
  pages={925--928},
  year={2003},
  publisher={Nature Publishing Group UK London}
}

@article{lee2012chemically,
  title={Chemically etched ultrahigh-{Q} wedge-resonator on a silicon chip},
  author={Lee, Hansuek and Chen, Tong and Li, Jiang and Yang, Ki Youl and Jeon, Seokmin and Painter, Oskar and Vahala, Kerry J},
  journal={Nature Photonics},
  volume={6},
  number={6},
  pages={369--373},
  year={2012},
  publisher={Nature Publishing Group UK London}
}

@article{yang2018bridging,
  title={Bridging ultrahigh-{Q} devices and photonic circuits},
  author={Yang, Ki Youl and Oh, Dong Yoon and Lee, Seung Hoon and Yang, Qi-Fan and Yi, Xu and Shen, Boqiang and Wang, Heming and Vahala, Kerry},
  journal={Nature Photonics},
  volume={12},
  number={5},
  pages={297--302},
  year={2018},
  publisher={Nature Publishing Group UK London}
}

@article{eichenfield2009picogram,
  title={A picogram-and nanometre-scale photonic-crystal optomechanical cavity},
  author={Eichenfield, Matt and Camacho, Ryan and Chan, Jasper and Vahala, Kerry J and Painter, Oskar},
  journal={Nature},
  volume={459},
  number={7246},
  pages={550--555},
  year={2009},
  publisher={Nature Publishing Group UK London}
}

@article{bereyhi2022hierarchical,
  title={Hierarchical tensile structures with ultralow mechanical dissipation},
  author={Bereyhi, Mohammad J and Beccari, Alberto and Groth, Robin and Fedorov, Sergey A and Arabmoheghi, Amirali and Kippenberg, Tobias J and Engelsen, Nils J},
  journal={Nature Communications},
  volume={13},
  number={1},
  pages={3097},
  year={2022},
  publisher={Nature Publishing Group UK London}
}

@article{fedorov2019generalized,
  title={Generalized dissipation dilution in strained mechanical resonators},
  author={Fedorov, Sergey A and Engelsen, Nils J and Ghadimi, Amir H and Bereyhi, Mohammad J and Schilling, Ryan and Wilson, Dalziel J and Kippenberg, Tobias J},
  journal={Physical Review B},
  volume={99},
  number={5},
  pages={054107},
  year={2019},
  publisher={APS}
}

@article{ghadimi2018elastic,
  title={Elastic strain engineering for ultralow mechanical dissipation},
  author={Ghadimi, Amir H and Fedorov, Sergey A and Engelsen, Nils J and Bereyhi, Mohammad J and Schilling, Ryan and Wilson, Dalziel J and Kippenberg, Tobias J},
  journal={Science},
  volume={360},
  number={6390},
  pages={764--768},
  year={2018},
  publisher={American Association for the Advancement of Science}
}

@article{engelsen2024ultrahigh,
  title={Ultrahigh-quality-factor micro-and nanomechanical resonators using dissipation dilution},
  author={Engelsen, Nils Johan and Beccari, Alberto and Kippenberg, Tobias Jan},
  journal={Nature Nanotechnology},
  volume={19},
  number={6},
  pages={725--737},
  year={2024},
  publisher={Nature Publishing Group UK London}
}

@article{beccari2022strained,
  title={Strained crystalline nanomechanical resonators with quality factors above 10 billion},
  author={Beccari, Alberto and Visani, Diego A and Fedorov, Sergey A and Bereyhi, Mohammad J and Boureau, Victor and Engelsen, Nils J and Kippenberg, Tobias J},
  journal={Nature Physics},
  volume={18},
  number={4},
  pages={436--441},
  year={2022},
  publisher={Nature Publishing Group UK London}
}

@article{huang2024room,
  title={Room-temperature quantum optomechanics using an ultralow noise cavity},
  author={Huang, Guanhao and Beccari, Alberto and Engelsen, Nils J and Kippenberg, Tobias J},
  journal={Nature},
  volume={626},
  number={7999},
  pages={512--516},
  year={2024},
  publisher={Nature Publishing Group UK London}
}

@article{ilgaz2024miniaturized,
  title={Miniaturized double-wing {$\Delta$E}-effect magnetic field sensors},
  author={Ilgaz, Fatih and Spetzler, Elizaveta and Wiegand, Patrick and Faupel, Franz and Rieger, Robert and McCord, Jeffrey and Spetzler, Benjamin},
  journal={Scientific Reports},
  volume={14},
  number={1},
  pages={11075},
  year={2024},
  publisher={Nature Publishing Group UK London}
}

@book{hetenyi1946beams,
  title={Beams on Elastic Foundation: Theory with Applications in the Fields of Civil and Mechanical Engineering},
  author={Het{\'e}nyi, M.},
  year={1946},
  publisher={University of Michigan Press},
  address = {Ann Arbor}
}

\end{document}